\documentclass[apj,iop]{emulateapj}
\usepackage{times}

\shorttitle{Contribution of Star-Forming Galaxies to EGRB}
\shortauthors{MAKIYA ET AL.}

\begin{document}

\title{Contribution from Star-Forming Galaxies to
                the Cosmic Gamma-Ray Background Radiation}

\author{Ryu Makiya and Tomonori Totani}
\affil{Department of Astronomy, Kyoto University,
     Kitashirakawa-Oiwake-cho, Sakyo-ku, Kyoto 606-8502, Japan}
\email{makiya@kusastro.kyoto-u.ac.jp}
\and
\author{Masakazu A. R. Kobayashi}
\affil{National Astronomical Observatory of Japan, 2-21-1 Osawa, Mitaka, Tokyo, 181-8588, Japan}

\begin{abstract}
  We present a new theoretical calculation of the contribution to the
  extragalactic gamma-ray background radiation (EGRB) from
  star-forming galaxies, based on a state-of-the-art model of
  hierarchical galaxy formation that is in quantitative agreement with
  a variety of observations of local and high-redshift galaxies.
  Gamma-ray luminosity ($L_\gamma$) and spectrum of galaxies are
  related to star formation rate ($\psi$), gas mass ($M_{\rm gas}$),
  and star formation mode (quiescent or starburst) of model galaxies
  using latest observed data of nearby galaxies.  We try the two
  limiting cases about gamma-ray production: the escape limit
  ($L_\gamma \propto \psi M_{\rm gas}$) and the calorimetric limit
  ($L_\gamma \propto \psi$), and our standard model predicts 7 and 4\%
  contribution from star-forming galaxies to the total EGRB flux
  (including bright resolved sources) recently reported by the {\it
    Fermi} Gamma-Ray Space Telescope. Systematic uncertainties do not
  allow us to determine the EGRB flux better than by a factor of $\sim$ 2.
  The predicted number of nearby galaxies detectable by {\it Fermi} is
  consistent with the observation. Intergalactic absorption by
  pair-production attenuates the EGRB flux only by a modest factor of
  $\sim$1.3 at the highest {\it Fermi} energy band, and the
  reprocessed cascade emission does not significantly alter EGRB at
  lower photon energies. The sum of the known contributions from AGNs
  and star-forming galaxies can explain a large part of EGRB, with a
  remarkable agreement between the predicted model spectrum and
  observation.
\end{abstract}


\section{INTRODUCTION}
\label{sec:INTRODUCTION}
The existence of the extragalactic diffuse gamma-ray background (EGRB)
has been revealed first by the SAS-2 satellite
(\citealt{1977ApJ...217L...9F}; \citealt*{1978ApJ...222..833F}).
Better determinations of the flux and spectrum of EGRB were achieved
by the EGRET detector on board the Compton Gamma-Ray
Observatory (\citealt{1998ApJ...494..523S};
\citealt*{2004ApJ...613..956S}). The most reliable measurement of EGRB
has very recently been reported based on the data of the {\it Fermi}
Gamma-Ray Space Telescope (\citealt{fermi}), and the EGRB spectrum is
well described by a single power-law with a photon index of $2.41 \pm
0.05$ and the photon flux of about $(1.03 \pm 0.17) \times 10^{-5}
\;{\rm photons\;cm^{-2}\;s^{-1}\;sr^{-1}}$ above 100 MeV
(\citealt{PhysRevLett.104.101101}).

The origin of EGRB has been discussed for a long time and various
sources have been discussed as possible contributors to EGRB, such as
active galactic nuclei (AGNs, especially blazars), galaxy clusters,
intergalactic shocks produced by structure formation, and dark matter
annihilation (see, e.g., \citealt{2007AIPC..921..122D} for a review).

Almost all of the known extragalactic gamma-ray sources are blazars,
and their contribution to the EGRB has been intensively studied
(e.g., Inoue \& Totani 2009; Venters 2010; Abdo et
al. 2010d, and references therein). 
However, star-forming galaxies should also be gamma-ray emitters by
cosmic-ray interactions with interstellar medium (ISM) and
interstellar radiation field (ISRF) (\citealt{2000ApJ...537..763S};
\citealt{2004ApJ...613..962S}), and there must be non-zero
contribution to EGRB. This is obvious because we know that the
Galactic disk is a strong source of diffuse gamma-rays, and gamma-rays
from Large Magellanic Cloud (LMC) have been detected by EGRET
(\citealt{1992ApJ...400L..67S}).  Furthermore, gamma-ray emission in
GeV--TeV from Small Magellanic Cloud (SMC,
  \citealt{2010arXiv1008.2127T})
and two nearby starburst galaxies,
M82 and NGC 253, have recently been discovered by
H.E.S.S. (\citealt{2009Sci...326.1080A}), VERITAS
(\citealt{2009Natur.462..770V}), and {\it Fermi}
(\citealt{2010ApJ...709L.152A}).  
The purpose of this paper is to
present a new estimate of the contribution from star-forming galaxies
to EGRB.

There are a number of previous studies on this issue (Strong et
al. 1976; Lichti et al. 1978; Dar \& Shaviv 1995; Pavlidou \& Fields
2002; Thompson et al. 2007; Bhattacharya \& Sreekumar 2009; Ando \&
Pavlidou 2009; Lacki et al. 2010; Fields et al. 2010), and the
estimates of the contribution to EGRB ranges $\sim$10--50\%\footnote{
  Here, the contribution is against the total (physical) extragalactic
  background photon flux ($>$100 MeV) including bright resolved
  sources. It is often calculated against the unresolved component of
  EGRB, but it depends on the flux sensitivity of a particular
  detector.  Throughout the paper, ``the contribution to EGRB'' is
  defined against the total EGRB flux, which we estimated from the
  resolved and unresolved components of the {\it Fermi} data (Abdo et
  al. 2010d).  We used a photon index of 2.4 for resolved blazars
  (Abdo et al. 2010d) to extrapolate the resolved component from 200
  to 100 MeV.}.  Most of these studies calculated gamma-ray luminosity
($L_\gamma$) from star formation rate (SFR) of galaxies, because the
cosmic-ray energy input is expected to be proportional to
SFR. However, if escape of cosmic-rays from galaxies is significant,
SFR cannot be used as a reliable indicator of $L_\gamma$ (see \S
\ref{sec:ModelingLg} in more details).  Several studies used infrared
luminosity of galaxies as SFR indicators, but IR luminosity
(especially at far-IR in early-type galaxies) is not a perfect SFR
estimator because a significant amount of dust can be heated by ISRF
from relatively old stars (e.g., Salim et al. 2009 and references
therein). 

Amount of interstellar gas should also be important to determine
$L_\gamma$, because the degree of cosmic-ray escape is affected by the
target amount.  In fact, recent observations indicate that gamma-ray
luminosity is nicely correlated with the product of SFR and gas mass
in galaxies(Abdo et al. 2010a).  Pavlidou \& Fields (2002)
included gas mass in the theoretical prediction of EGRB using the data
of the cosmic star formation history (CSFH). 
Galaxies in the universe is considered as a closed box containing
stellar and gas mass in present-day galaxies, and CSFH is used to
solve the time evolution of relative fractions of stars and gas.  
However this approach likely overestimates the
gas mass contributing to gamma-ray production at high redshifts,
because most of the present-day stellar mass is in the form of gas and
assumed to contribute to gamma-ray emission.  In reality, only the gas
in collapsed dark halos can contribute to cosmic-ray interactions in
galaxies, which is a small fraction at high redshifts according to the
structure formation theory. Recently, Fields et al. (2010)
incorporated gas mass in galaxies using the Schmidt-Kennicutt relation
in their calculation of EGRB.  In this case one must know galaxy size
to relate SFR and $M_{\rm gas}$, and a single value of galaxy size at
each redshift was assumed as a simple model.

Here we present a new calculation of the contribution from
star-forming galaxies in EGRB using a state-of-the-art theoretical
model of galaxy formation in the framework of hierarchical galaxy
formation, which is in quantitative agreement with a variety of
observations at high redshifts as well as the local universe. An
advantage of our approach compared with previous studies is that we
can calculate both SFR and gas mass of individual galaxies at various
redshits.  Furthermore, we utilize the information of the gamma-ray
spectra recently observed for nearby starbursts, in addition to the
standard spectrum of the Galactic diffuse emission, to predict the
EGRB spectrum based on the galaxy formation model including both
quiescent and starburst galaxy populations.

The paper is organized as follows. In the next section we describe our
model calculations. We present the results on EGRB and statistics
about the number of nearby galaxies detectable by {\it Fermi} in \S
\ref{sec:RESULTS}. After discussion on the uncertainties in our
calculation (\S \ref{sec:Uncertainties}) and implications for the
origin of EGRB (\S \ref{sec:EGRB-origin}), conclusions will be
presented in \S \ref{sec:Conclusions}.  In this work, cosmological
parameters of $\Omega_{0}=0.3$, $\Omega_{\Lambda}=0.7$, and
$H_{0}=70~{\rm Mpc^{-1}~km~s^{-1}}$ are adopted.

\section{Formulations}
\label{sec:Formulations}
\subsection{The Theoretical Model of Hierarchical Galaxy Formation}
\label{sec:Mitakamodel}
We use a mock numerical galaxy catalog produced by one of the latest
so-called semi-analytic models (SAMs) of hierarchical galaxy formation
[the Mitaka model (\citealt{2004ApJ...610...23N};
\citealt{2005ApJ...634...26N})].  In general, SAMs compute merging
history of dark matter (DM) halos based on the standard structure
formation theory driven by cold dark matter, and include several
important physical processes related to the evolution of baryons in DM
halos such as radiative gas cooling, star formation, supernova
feedback, galaxy merger, stellar population synthesis, chemical
evolution, and extinction by interstellar dust
 (e.g., \citealt{1993MNRAS.264..201K};
\citealt{1994MNRAS.271..781C}; \citealt{1999MNRAS.305..449N};
\citealt{1999MNRAS.310.1087S}; \citealt{2005MNRAS.356.1191B}).  In
each time step of the halo merging history, two discrete modes of star
formation are considered in the model, i.e., starburst and quiescent.
When galaxies experience a major galaxy merger, intensive burst of
star formation is assumed to occur (the starburst mode), and all of
the available cold gas is converted into stars and hot gas within a
short time. Otherwise star formation proceeds at a modest rate
determined by cold gas amount and dynamical time scales (the quiescent
mode).

The Mitaka model can quantitatively reproduce a wide variety of
observed characteristics of local galaxies, including luminosity
functions (LFs) and scaling relations among various observables such
as magnitude, colors, surface brightness, size, gas mass-to-light ratio, and
metallicity (Nagashima \& Yoshii 2004).  Moreover, it can also
reproduce well the cosmic star formation history (Nagashima \& Yoshii
2004), the rest-frame ultraviolet (UV) continuum LF of Lyman break
galaxies at $z=4$ and 5 (\citealt{2006ApJ...648....7K}), and all of
the available observations for the high-$z$ Ly$\alpha$ emitters
(Ly$\alpha$ and UV continuum LFs, and Ly$\alpha$ equivalent width
distributions) (\citealt{2007ApJ...670..919K},
\citeyear{2010ApJ...708.1119K}).

In the version of the Mitaka model used here, about 100 Monte-Carlo
realizations of merger histories are produced for DM halos in a DM
mass interval of 0.1 dex at each output redshift ranging $z=$
0--10. The model calculates DM halos with velocity dispersions larger
than 30 km/s, and this allows us to make a reliable prediction in the
stellar mass range of $\gtrsim 10^6 M_\odot$, roughly corresponding to
the absolute $B$ band magnitude of $M_B \lesssim -10$ mag.

\subsection{Modeling Gamma-Ray Emission from Galaxies}
\label{sec:ModelingLg}

To calculate the EGRB flux from normal galaxies, we need the gamma-ray
LF $\phi (L_{\gamma }, z)$ (comoving number density of galaxies per
unit gamma-ray luminosity) as a function of redshift. Here, $L_\gamma$
is defined as gamma-ray luminosity in the rest-frame photon energy range of
0.1--5 GeV.  The LF $\phi$ is calculated from the mock galaxy catalog
by modeling gamma-ray luminosity of each galaxy as follows.

Diffuse gamma-ray radiation in galaxies is produced by the
interactions between the cosmic-rays and ISM. If the cosmic-ray losses
are dominated by escape, the gamma-ray luminosity is set by
equilibrium between cosmic ray production and escape.  It is then
expected that gamma-ray luminosity depends on SFR ($\psi$, an
indicator of cosmic ray production rate) and mass of interstellar gas
($M_{\rm gas}$, the amount of target for cosmic rays to interact
before escape), i.e., $L_\gamma \propto \psi M_{\rm gas}$.  We call it
``the escape limit''.  However, if the cosmic-ray energy losses are
dominated by inelastic collision with interstellar gas and almost all
of the cosmic-ray energy is converted into gamma-rays, the gamma-ray
luminosity is no longer dependent on the amount of gas, i.e.,
$L_\gamma \propto \psi$.  We call it ``the calorimetric limit''. [See
also Pavlidou \& Fields (2001), Torres et al. (2004), and Thompson et al.(2007)
 for more discussion on this issue.]  Since it is difficult to model the
detailed escape and energy loss processes of cosmic-rays in galaxies,
we consider both of these two simple limiting cases.

To establish the link between $L_\gamma$ and $\psi$ and/or $M_{\rm
  gas}$, we collected the observed values of these quantities for the
nearby galaxies of SMC, LMC, the Milky Way (MW), M82, and NGC 253. They
are summarized in Table \ref{table:Lg-SFR-Mgas}. For SMC and LMC, SFRs
are estimated from H$\alpha$ luminosities \citep{2008ApJS..178..247K},
using the relationship of \cite{1998ApJ...498..541K} assuming the
Salpeter initial mass function (IMF) in 0.1--100 $M_\odot$. Extinction
by dust is not taken into account, and the uncertainty about this will
be discussed later (\S \ref{sec:Uncertainties}). For M82 and NGC 253,
we used SFR estimated from the IR luminosity from interstellar dust,
using the relationship of \cite{1998ApJ...498..541K} assuming the same
IMF as above, since these galaxies has large extinction
(\citealt{2007ApJ...655..863D} and \citealt{1980ApJ...239...54P}) and
H$\alpha$ is no longer a good indicator of SFR. Total IR luminosity is
calculated from the IRAS fluxes in the three bands
(\citealt{2003AJ....126.1607S}) using the formula given in
\cite{2002ApJ...576..159D}.  Since it is difficult to estimate the
total IR luminosity of MW, we use the supernova rate estimated by a
gamma-ray observation (\citealt{2006Natur.439...45D}) and converted it
into SFR assuming the same IMF.  The standard value of 8 $M_\odot$ is
adopted for the mass threshold of core-collapse supernova explosions.

The gas mass is estimated by the total of ${\rm H_{2}}$ and
\ion{H}{1}, where the former and the latter are measured by CO and
21-cm line observations, respectively.  The references are shown in
Table \ref{table:Lg-SFR-Mgas}. There are various estimations for
\ion{H}{1} mass of SMC and LMC by different authors
(\citealt{1989gcho.book.....H}, \citealt{1999MNRAS.302..417S},
\citealt{2005A&A...432...45B} for SMC, and
\citealt{1989gcho.book.....H}, \citealt{1992A&A...263...41L},
\citealt{1997macl.book.....W}, \citealt{2003MNRAS.339...87S},
\citealt{2005A&A...432...45B} for LMC).  Here we adopted the latest
value by \cite{2005A&A...432...45B}, and the standard deviations of
the \ion{H}{1} mass measurements among difference papers were added to
the errors as a quadratic sum, to take into account the uncertainty
(Table \ref{table:Lg-SFR-Mgas}).

\begin{figure}
	\epsscale{1.2}
	\plotone{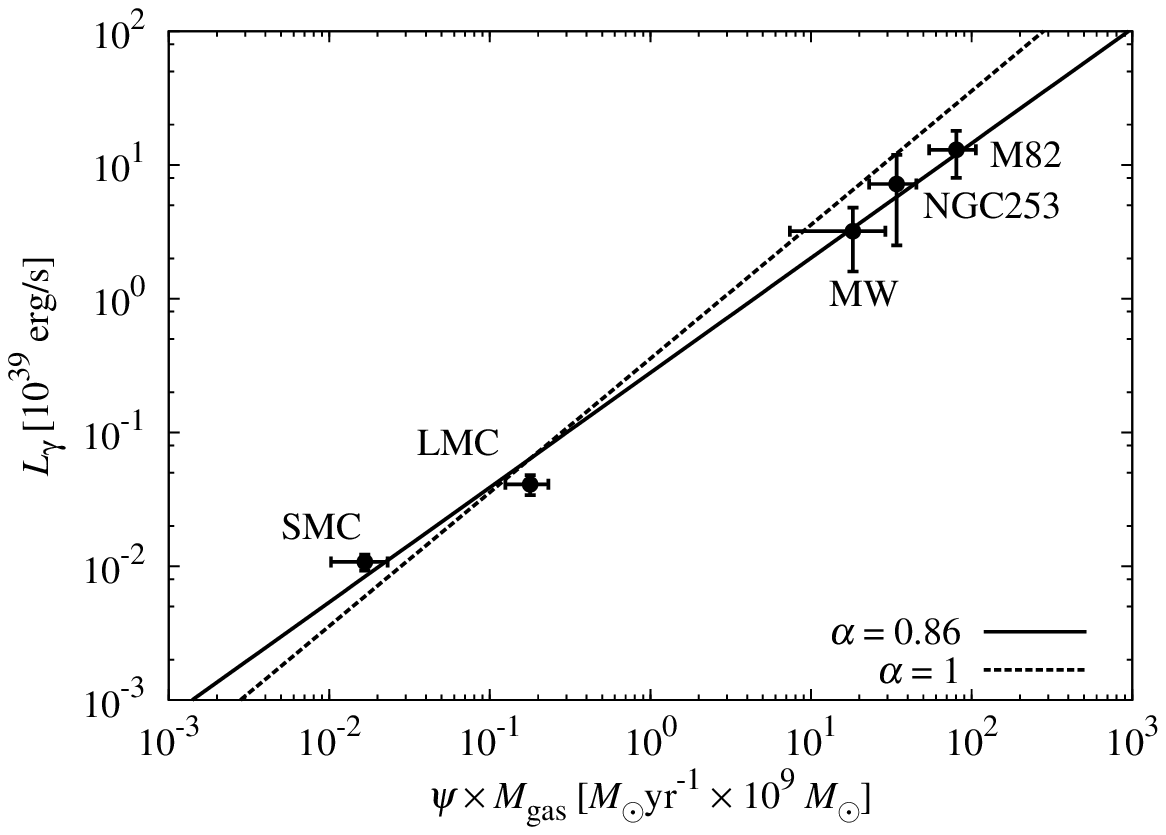}
	\plotone{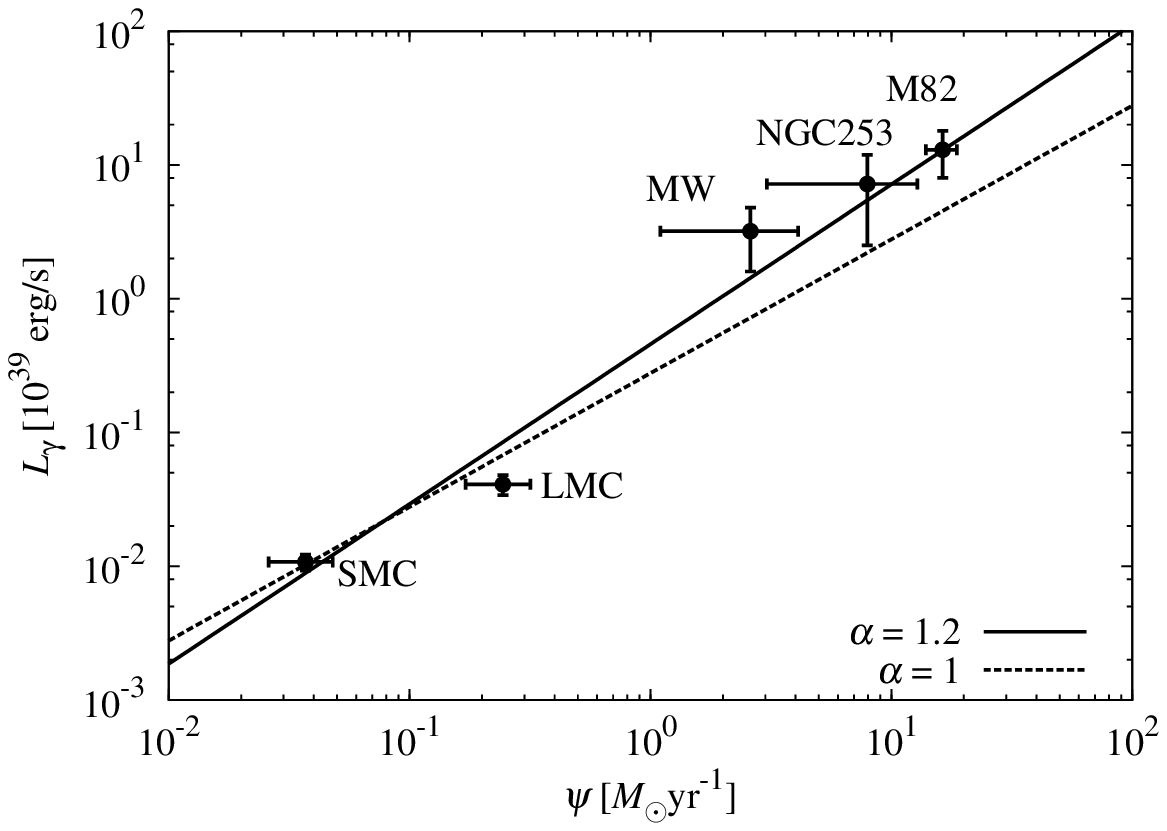}
	\caption{({\it Top}) Correlation between the gamma-ray
          luminosity $L_\gamma$ (0.1--5 GeV) and the product of SFR
          and gas mass, $\psi M_{\rm gas}$. The solid line represents
          the best fit power-law $L_\gamma \propto (\psi M_{\rm
            gas})^\alpha$ with $\alpha$ taken as a free parameter.
          The dashed line represents a linear function fit with a
          fixed value of $\alpha = 1$. ({\it Bottom}) The same as top
          panel, but for the correlation between $L_\gamma$ and SFR.}
\label{fig:correlation}
\end{figure}

The collected data are plotted in Figure \ref{fig:correlation}.  It is
clearly seen that there is a good correlation both for the
$L_\gamma$-$\psi M_{\rm gas}$ (top panel) and $L_\gamma$-$\psi$ relations
(bottom panel).  We fitted these relations by a single power-law
function to the data points of SMC, LMC, M82, and NGC 253, and the
results are
\begin{eqnarray}
L_{\gamma} &=& (0.28 \pm 0.07)  \nonumber \\
&\times& \left( \frac{\psi}{M_{\odot }\; {\rm yr}^{-1}} 
\times \frac{M_{\rm gas}}{10^{9}M_{\odot }} \right)^{0.86 \pm 0.06} 
[10^{39} {\rm erg/s}] \ 
\label{eq:Lg-SFR-Mgas}
\end{eqnarray}
and
\begin{eqnarray}
L_{\gamma} &=& (0.46 \pm 0.12)  \nonumber \\
&\times& \left( \frac{\psi}{M_{\odot }\; 
{\rm yr}^{-1}} \right )^{1.2 \pm 0.09} 
[10^{39} {\rm erg/s}] \ ,
\label{eq:Lg-SFR}
\end{eqnarray}
where the errors are statistical 1$\sigma$. In the fitting calculation
we adopted the effective variance method which takes into account the errors
of both coordinates (Orear 1981).
Since SFR of MW has a large uncertainty, we did not use the data of MW in 
the fitting, although it is consistent with the correlation.  The exact 
proportionality between $L_\gamma$ and $\psi M_{\rm gas}$ or $L_\gamma$ 
and $\psi$ are not necessarily expected, because it depends on propagation 
and degree of confinement of cosmic rays within a galaxy. Therefore we 
use these power-law relation to calculate gamma-ray luminosity.  
The cases of the exact proportionality, i.e., $L_\gamma \propto (\psi M_{\rm
  gas})^1$ and $L_\gamma \propto (\psi)^1$ will also be discussed
later (\S \ref{sec:Uncertainties}).

It should be noted that $M_{\rm gas}$ in the Mitaka model has been
compared with observations of local galaxies as a function of $B$ band
galaxy luminosity (Nagashima \& Yoshii 2004; Nagashima et al. 2005),
showing a good agreement.  This gives a support to apply $M_{\rm gas}$
of the Mitaka model to the $L_{\gamma}$-$\psi M_{\rm gas}$ relation
determined by observations, though the model has not yet been tested
against $M_{\rm gas}$ at high redshifts because only very few
observations are available.  
As for SFR, the adopted IMF in the
Mitaka model (the Salpeter in 0.1--60 $M_\odot$) is slightly different
from that (in 0.1--100 $M_\odot$) assumed in converting the observed
IR luminosity to SFR in Fig. \ref{fig:correlation}. Though the value
of SFR is hardly affected by the upper bound of stellar mass in IMF,
it may affect the conversion from IR luminosity to SFR.  However we
confirmed, using the PEGASE stellar library (Fioc \& Rocca-Volmerange
1997), that the difference of the IMF mass ranges hardly affects the
total luminosity from young stars because stars heavier than 60
$M_\odot$ are scarce and have short lifetimes. It is negligible at the
stellar population ages larger than $10^{6.5}$ yrs, and at most 20\%
at ages younger than that. Therefore we can apply SFRs in the Mitaka
model to the fitting formula eqs. (\ref{eq:Lg-SFR-Mgas}) and
(\ref{eq:Lg-SFR}).

\subsection{Gamma-Ray Spectra of Galaxies}
\label{sec:spectra}

\begin{figure}
	\epsscale{1.2}
	\plotone{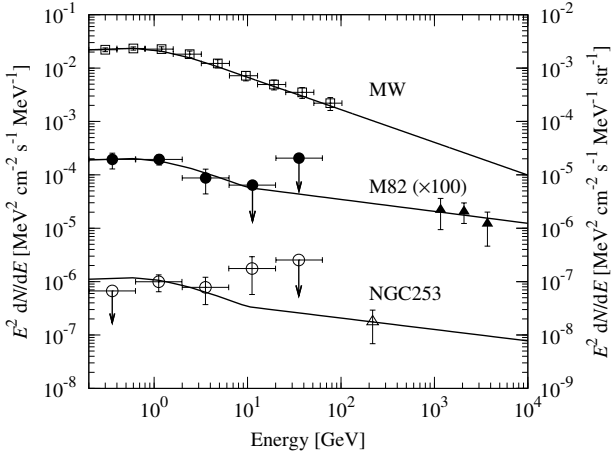}
	\caption{ The gamma-ray spectrum $E^2 dN/dE$ of MW, M82 and
          NGC 253, where $E$ is gamma-ray energy and $dN/dE$ the
          differential photon flux.  The MW spectrum is diffuse
          background flux in units shown on the right ordinate, and the
          others are flux from a source in units shown on the left
          ordinate.  For the purpose of presentation, the spectrum and
          the data points of M82 are multiplied by 100. The MW data
          (open squares) are the Galactic diffuse gamma-rays averaged
          over $|b|>10^{\circ }$ (\citealt{PhysRevLett.104.101101}).
          The {\it Fermi} data of M82 (filled circles) and NGC 253
          (open circles) are obtained by
          \cite{2010ApJ...709L.152A}. The data in the very high-energy
          range of M82 (filled triangles,
          \citealt{2009Natur.462..770V}) and NGC 253 (open triangles,
          \citealt{2009Sci...326.1080A}) are also plotted.  The solid
          curves are the model spectra used to calculate EGRB (see
          text for details).}
\label{fig:specs}
\end{figure}

We also need to model the gamma-ray spectrum of galaxies to predict
the EGRB spectrum. The observed gamma-ray spectra of MW, M82, and NGC
253 are shown in Figure \ref{fig:specs}.  We simply assume that
quiescent galaxies have the same spectral shape as that of MW. For the
MW spectrum we used the GALPROP fit to the {\it Fermi} spectral data
of the Galactic diffuse gamma-rays (\citealt{PhysRevLett.104.101101}).
The observed spectrum of SMC and LMC are similar to that of MW.  

For starburst galaxies, we expect some change of gamma-ray spectrum by
the following physical reason. Because of the high density of star
forming regions, it is plausible that starburst galaxies are
``calorimetric'' and have harder spectrum than quiescent galaxies
because of less significant espace of high energy cosmic rays.
Therefore we apply two modelings for the starburst spectra: (1)
starbursts have the same spectrum as MW, and (2) the MW-like spectrum
at low energy range and a hard power-law tail at high energy range.
We fitted the spectrum of M82 and NGC 253 by the MW spectrum at energy
less than 10 GeV, and with power law above the 10 GeV. Normalization
of the power law component was determined so that it smoothly connects
to the MW spectrum at 10 GeV.  (Therefore the free parameters of the
fit are two: the normalization of MW component and the photon index of
high-energy power-law component.)  As a result of the fitting, we find
that M82 and NGC 253 can be fit with almost the same photon index of
2.2, and hence we adopt this spectrum when we try a different spectrum
from MW for starburst galaxies\footnote{In the spectral fitting, we
  ignored the data points that have only upper limits.}.  These model
spectra are shown in Figure \ref{fig:specs}.

\subsection{The EGRB Calculation}
\label{sec:EGRBcalc}
\begin{figure}
	\epsscale{1.2}
	\plotone{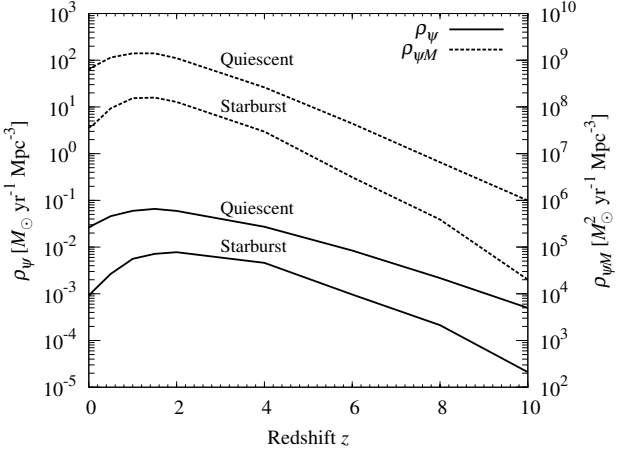}
        \caption{The evolution of the comoving densities of SFR
          ($\rho_\psi$, solid line) and the product of SFR and gas
          mass ($\rho_{\psi M}$, dashed line) in galaxies, calculated
          by the model of hierarchical galaxy formation.  See the left
          and right ordinates for the units of $\rho_{\psi}$ and
          $\rho_{\psi M}$, respectively.
          }
\label{fig:csfh}
\end{figure}

Given the modeling of gamma-ray luminosity of galaxies described
above, the EGRB flux and spectrum can be calculated by integrating the
gamma-ray luminosity function $\phi(L_\gamma, z)$ over redshift.  To
see the redshift evolution of the gamma-ray emissivity, we show the
redshift evolution of $\rho_\psi$ (comoving SFR density, proportional
to gamma-ray luminosity density ${\cal L_\gamma}$ in the calorimetric
limit) as well as $\rho_{\psi M}$ (the comoving density of $\psi
M_{\rm gas}$, proportional to ${\cal L_\gamma}$ in the escape
limit) in Figure \ref{fig:csfh}.  This figure indicates that
$\rho_{\psi M}$ decreases more rapidly toward higher redshift than
$\rho_{\psi}$, and this is likely because $\psi$ is roughly
proportional to $M_{\rm gas}$ and hence both the cosmic densities of
$\psi$ and $M_{\rm gas}$ decrease to high redshift beyond $z \sim 1$.

Given the gamma-ray luminosity function of galaxies $\phi(L_\gamma, z)$,
the EGRB spectrum (photon flux per unit photon energy per steradian)
is expressed as 
\begin{eqnarray}
\frac{d^{2}F(\epsilon_{\gamma })}{d\epsilon_{\gamma }d\Omega } 
&=& \frac{c}{4\pi } \int^{z_{\max}}_{0} \left|\frac{dt}{dz}\right| dz 
\int^{\infty }_{0} dL_{\gamma } \phi (L_{\gamma }, z) \nonumber \\
&\times& (1+z)
\frac{dL_\gamma [L_{\gamma }, (1+z)\epsilon_{\gamma }]}{d\epsilon_{\gamma,r}} 
e^{-\tau_{\gamma }(z, \epsilon_{\gamma })} \ ,
\end{eqnarray}
where $\epsilon_{\gamma }$ is photon energy for observers at $z=0$,
$t$ the cosmic time, and $dt/dz$ can be calculated if the standard
cosmological parameters are given.  The rest-frame gamma-ray spectrum
of galaxies $dL_\gamma(L_\gamma, \epsilon_{\gamma, r})
/d\epsilon_{\gamma, r}$ can be calculated for a given $L_\gamma$ by
the spectral shapes as modeled in the previous section, where
$\epsilon_{\gamma,r }$ is a photon energy at the rest-frame of the
source redshifts. We set the maximum redshift $z_{\rm max} = 10$
corresponding to that of the galaxy formation model, but this
parameter hardly affects the EGRB flux because the majority of the EGRB
flux comes from galaxies at $z \lesssim 2$ 
(see \S \ref{sec:RESULTS} below).

High energy photons ($\gtrsim $ 20 GeV) are absorbed during the
propagation in intergalactic space, by interaction with the cosmic
infrared background (CIB) photons producing electron-positron pairs
(\citealt{1998ApJ...493..547S}; \citealt{2002ApJ...570..470T};
\citealt{2004A&A...413..807K}; \citealt{2006ApJ...648..774S}).  The
optical depth $\tau_{\gamma }(z, \epsilon_{\gamma })$ in equation (2)
represents this interaction. Since the infrared emission from dust has
not yet been included in the Mitaka model, $\tau_\gamma$ cannot be
calculated self-consistently based on the Mitaka model.  Therefore we
use the model of \cite{2002ApJ...570..470T} for $\tau_{\gamma }$ in
this work, which is based on a different galaxy evolution model. To
examine the effect of the uncertainty about $\tau_{\gamma }$, we also
calculated with different models of $\tau_\gamma$ by
\cite{2004A&A...413..807K} and \cite{2008A&A...487..837F}, and found
that the change in the EGRB flux is not significant (negligible at
$\lesssim$ 30 GeV, and at most by a factor of 1.5 at the highest {\it
  Fermi} energy band).

The created pairs would scatter up the
cosmic microwave background photons to gamma-rays via inverse Compton
mechanism (cascade emission) and they make some contributions to the
EGRB flux at lower photon energies (\citealt{1994ApJ...423L...5A};
\citealt{2004A&A...415..483F}).  We calculate the contribution from
the cascade emission using the same formalism of
\cite{2008A&A...479...41K}.

\begin{table}
\begin{center}
\caption{Summary of $L_{\gamma }$, ${\rm SFR}$, ${\rm {\it M}_{gas}}$, and Distance of Nearby Galaxies \label{table:Lg-SFR-Mgas}}
\begin{tabular}{ccccc}
\tableline\tableline
Object & $L_{\gamma}$\tablenotemark{a} & SFR\tablenotemark{b} & $M_{{\rm gas}}$\tablenotemark{c} & Distance\tablenotemark{d}\\
{}  & [$10^{39}$ erg/s] & ${\rm [M_{\odot}/yr]}$ & [$10^{9}{\rm M_{\odot}}$] & [Mpc] \\
\tableline
SMC & $0.011 \pm 0.001$ & $0.037 \pm 0.011$ & $0.45 \pm 0.11$ & $0.061 \pm 0.003$\\
LMC & $0.041 \pm 0.007$ & $0.244 \pm 0.073$ & $0.56 \pm 0.14$ & $0.049 \pm 0.001$\\
M82 & $13.0 \pm 5.0$ & $16.3 \pm 2.4$ & $4.9 \pm 0.58$ & $3.6 \pm 0.3$ \\
NGC 253 & $7.2 \pm 4.7$ & $7.9 \pm 4.9$ & $4.3 \pm 0.55$ & $3.9 \pm 0.4$ \\
MW & $3.2 \pm 1.6$ & $2.6 \pm 1.5$ & $7.0 \pm 1.0$ & -- \\
\tableline
\end{tabular}
\footnotetext{Gamma-ray luminosities are in 0.1--5 GeV, from \cite{2010arXiv1008.2127T} for SMC, 
		        \cite{2010A&A...512A...7A} for LMC, and \cite{2010ApJ...709L.152A} for M82, NGC 253 and MW.}
\footnotetext{See text (\S \ref{sec:ModelingLg}) for methods and references of SFR estimates.}
\footnotetext{Total gas mass (\ion{H}{1} + $\rm H_2$) are estimated from 21-cm and CO line fluxes. 
        References are \cite{2005A&A...438..533W}, \cite{1989gcho.book.....H} (for M82), 
        \cite{2007ApJ...666..156K}, \cite{2005ApJS..160..149S} (for NGC 253), 
        and \cite{1999MNRAS.307..857B} (for MW).
        For SMC and LMC, see text (\S \ref{sec:ModelingLg}) for methods and references of 
		gas mass estimates.}
\footnotetext{References are \cite{2005MNRAS.357..304H} for SMC, \cite{2009ApJ...697..862P} for LMC, 
		\cite{1994ApJ...427..628F} for M82, and \cite{2003A&A...404...93K} for NGC 253.}
\end{center}
\end{table}

\section{RESULTS}
\label{sec:RESULTS}
\subsection{EGRB Predictions Compared with Observation}
\label{sec:EGRBwithObs}

Figure \ref{fig:egrb1} shows the quiescent and starburst components of
EGRB calculated by the formulations presented in the previous section,
for the two different modelings of the escape and calorimetric limits
in different two panels.  The data points are the {\it Fermi}
measurements of EGRB flux (\citealt{PhysRevLett.104.101101}).  The
reported {\it Fermi} EGRB spectrum is the residual after the sources
resolved by {\it Fermi} have been subtracted, and it is difficult to
compare a theoretical prediction to such data taking into account
complicated detection process and efficiencies.  Therefore we also
plot the total {\it Fermi} EGRB flux including the contribution from
resolved sources as reported by \cite{PhysRevLett.104.101101}, and
compare our prediction including all sources in the universe with this
total observed flux.  For the starburst galaxies we plotted the two
models using the two different spectral shapes (MW or MW$+$power-law)
as discussed in \S \ref{sec:spectra}.  All the predictions take into
account the intergalactic absorption and the reprocessed cascade
emission as discussed in the previous section.

In Figure \ref{fig:egrb2}, the predictions of total EGRB flux from all
star-forming galaxies (quiescent plus starburst) are plotted.  To show
the effect of intergalactic absorption, the EGRB spectrum without
taking into account the absorption is also plotted. The cascade
component produced by the absorption is also shown.  It can be seen
that the intergalactic absorption attenuates the EGRB flux by a factor
of about 1.3 at the highest photon energy of {\it Fermi}.  The
fraction of the absorbed energy flux of EGRB is not a large amount
compared with the total EGRB energy flux, and hence the cascade
component does not make a significant contribuiton at the low energy
bands (Note that Fig. \ref{fig:egrb2} is made as a $\nu F_\nu$ plot).

The EGRB flux from starburst galaxies is significantly enhanced at
high energy range of $\epsilon_\gamma \gtrsim 10$ GeV when the
MW$+$power-law spectra is applied. However, the difference is not
significant in the total EGRB flux from all star-forming galaxies,
provided that quiescent galaxies have similar gamma-ray spectra to
that of MW.  It should also be noted that the classification of
quiescent and starburst galaxies in the Mitaka model is determined to
reproduce the optical/infrared observed properties of galaxies. 
Therefore the classification can be systematically
different if we do it by the difference of gamma-ray spectra. This is
an interesting question but difficult to answer at the present time
because of too few observed data of gamma-ray spectra of galaxies.

The EGRB photon flux from all star-forming galaxies is 9.9 $ \times
10^{-7}$ (escape limit) and 5.6 $\times 10^{-7}$
(calorimetric limit) ${\rm photons\;cm^{-2}\;s^{-1}\;sr^{-1}}$ ($>$
100 MeV), which are 7.0$\pm1.8$\% and 3.9$\pm1.0$\% of the EGRB flux
observed by {\it Fermi} (including resolved sources),
respectively.  
If we compare these results with 
the recent {\it Fermi} measurement of the unresolved EGRB flux, 
the contribution from star-forming galaxies
become 9.6$\pm2.4$\% (escape limit) and 5.4$\pm1.4$\% (calorimetric limit).
Here the error is that coming from the statistical
1$\sigma$ error of the fitting formula of $L_\gamma$
[eqs. (\ref{eq:Lg-SFR-Mgas}) and (\ref{eq:Lg-SFR})].

As a result of calculation the EGRB flux from star-forming galaxies in
the calorimetric limit becomes lower than the escape limit.  This
seems counter-intuitive, since more gamma-rays should be produced in
the calorimetric limit if the cosmic-ray energy input is fixed.  This
is because we use the datapoints of SMC and LMC to derive
eq. (\ref{eq:Lg-SFR}). These galaxies are likely in the escape limit,
making the slope of $\psi$-$L_\gamma$ relation steeper. It would be an
overestimate of EGRB if we calculate the ``truly'' calorimetic case,
i.e., $L_\gamma \propto \psi^1$ with the proportionality constant for
starburst galaxies. Rather, what we calculated here is the best
empirical estimate of EGRB in a model in which $L_\gamma$ is a
function of $\psi$.

\subsection{Redshift Distribution of EGRB Photons}

\begin{figure}
	\epsscale{1.2}
	\plotone{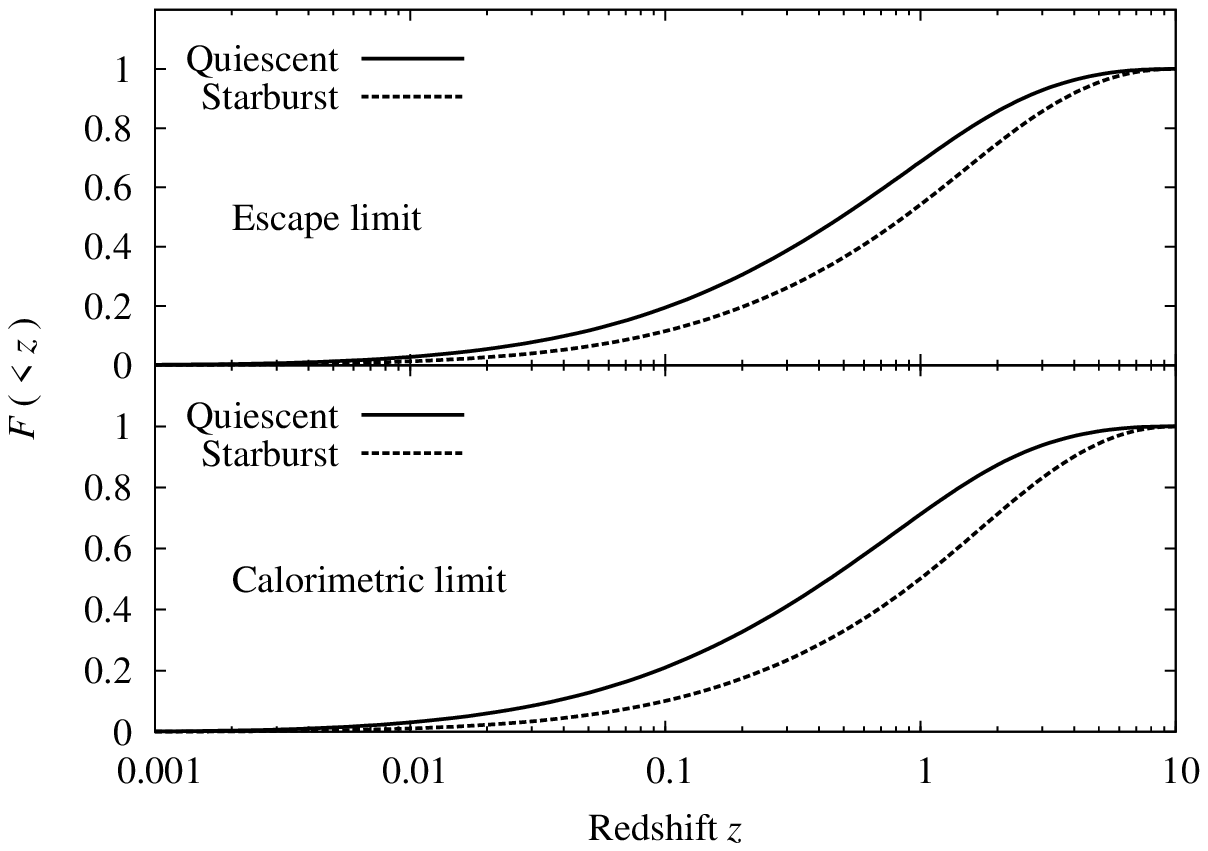}
	\caption{Cumulative redshift distribution of the EGRB photons
          (0.1--5 GeV) from star-forming galaxies. Quiescent
          and starburst galaxies are separately shown by solid and
          dashed curves, respectively.}
\label{fig:fz}
\end{figure}

Figure \ref{fig:fz} shows the cumulative redshift distribution of the
EGRB photons from star-forming galaxies.  It is clearly seen in the
figure that more than half of the total EGRB flux comes from galaxies
at $z < 1$, and more than 90\% from $z < 2$ both for the escape- and
calorimetric-limit models.  This is a reasonable result given the
evolution of $\rho_{\psi}$ and $\rho_{\psi M}$ as discussed in
\S\ref{sec:EGRBcalc}. The distribution of photons from starburst
galaxies extend to slightly larger redshifts than that from quiescent
galaxies, because of the stronger evolution of $\rho_{\psi M}$ at $z
\sim $ 0--1. Our result of the escape-limit can be compared with that
of Ando \& Pavlidou (2009, hereafter AP09). The two models are in
rough agreement, though the median redshift of our model is slightly
smaller than that of AP09. This may be because AP09 assumes
that all the mass in the present-day stars 
contributes to gamma-ray production as gas in the early universe,
while only a small fraction is in the form of cooled gas in our
model, as discussed in \S \ref{sec:INTRODUCTION}.

\subsection{Detectability of Nearby Star-Forming Galaxies}
\label{sec:Detectability}
We also calculated the number of galaxies that are detectable by {\it
  Fermi}. The 1-yr sensitivity of {\it Fermi} is $\sim 3 \times
10^{-9} \ \rm photons \ cm^{-2} s^{-1}$ at $\epsilon_\gamma > 100$
MeV (\citealt{fermi}). Our model predicts
4.6 and 2.3 galaxies above this flux in the
escape- and calorimetric-limit models, respectively. These numbers are
consistent with the actual number of four (SMC, LMC, M82, and NGC 253)
or two (when SMC and LMC are excluded because it is within our Galaxy
halo; see below).  This result gives another support to the
reliability of our EGRB prediction. However, it should be noted that
in this estimate all galaxies in the Mitaka model are assumed to
distribute randomly in space with a uniform mean density, because the
clustering information of galaxy distribution is not included in the
semi-analytic models like the Mitaka model. The density fluctuation
and clustering in the local group around MW should affect the above
estimate, especially for satellite galaxies like LMC. Not only the
clustering effect but also the realistic detection efficiency of {\it
  Fermi} sources around the detection limit must be properly taken
into account in a more quantitative analysis, which is beyond the
scope of this paper.  In the future, {\it Fermi} sensitivity will
reach $\sim 1 \times 10^{-9} {\ \rm photons\;cm^{-2}\;s^{-1}}$ ($>$
100 MeV) by the 10-year observation, and we expect that {\it Fermi}
will detect $\sim$ 24 (escape limit) or 12
(calorimetric limit) nearby star-forming galaxies, though it is
again subject to the uncertainty about clustring, and is probably
optimistic assuming 100\% detection efficiency above that flux.

Since the {\it Fermi} sensitivity becomes worse for extended sources
than that for point sources, we checked the size of galaxies computed
by the Mitaka model. There is a large scatter in the correlation
between size and gamma-ray luminosity, but the average angular size of
model galaxies brighter than {\it Fermi} 1-yr sensitivity is
$\sim$0.05 deg, which is much smaller than the {\it Fermi} angular
resolution ($\sim0.6$ deg at 1 GeV, \citealt{fermi}). Therefore the
effect of extended sources should not significantly affect the above
estimate for the number of detectable sources. It should also be noted
that the size of an active star-forming region in a galaxy can be much
smaller than that of the stellar disk of the same galaxy, as inferred
from observations of NGC 253 (\citealt{2009Sci...326.1080A}).
Therefore gamma-ray emitting regions can be much smaller than the size
of the model galaxies in the Mitaka model.

\section{Discussion}
\subsection{Uncertainties}
\label{sec:Uncertainties}

Here we examine the uncertainties about our estimate of the
star-forming galaxy contribution to EGRB.  As described in \S
\ref{sec:ModelingLg}, we have calculated gamma-ray luminosity of
galaxies by the relations $L_\gamma \propto (\psi M_{\rm gas})^{0.86}$
and $L_\gamma \propto (\psi)^{1.2}$ which are obtained by the fit to
the data of SMC, LMC, M82, and NGC 253. We also tested the cases of
$L_\gamma \propto (\psi M_{\rm gas})^1$ and $L_\gamma \propto
(\psi)^1$, and in this case, the best-fit relations to the data
becomes
\begin{eqnarray}
L_{\gamma} &=& (0.36\pm 0.08) \nonumber \\
&\times& \left( \frac{\psi}{M_{\odot }\; {\rm yr}^{-1}} \times 
\frac{M_{\rm gas}}{10^{9}\;M_{\odot }} \right) [10^{39} {\rm erg/s}],
\label{eq:Lg-SFR-Mgas-index1}
\end{eqnarray}
and
\begin{equation}
L_{\gamma} = (0.28 \pm 0.08) \times \left( 
\frac{\psi}{M_{\odot }\; {\rm yr}^{-1}} \right) [10^{39} {\rm erg/s}] .
\end{equation}
The eq. (\ref{eq:Lg-SFR-Mgas-index1}) is consistent with the estimate
of the same relation by \cite{2001ApJ...558...63P} within the
uncertainty of the conversion of MW supernova rate into SFR. We found
that in these cases the EGRB flux becomes slightly different from our
standard models: 9.3$\pm$2.0\% (escape limit) and 1.8$\pm$ 0.5\%
(calorimetric limit) contribution to the observed total {\it Fermi}
EGRB flux (or 12.8$\pm2.8$\% and 2.5$\pm0.7$\% against 
the unresolved component of EGRB reported by {\it Fermi}). 
The value of 12.8$\pm2.8$\% can be compared with another recent
result by Fields et al. (2010), based on more empirical calculations
about galaxy eovlution, and they are consistent with each other
within the systematic uncertainties. 

Estimate of SFR is generally affected by the treatment of extinction
by interstellar dust.  The SFRs of LMC and SMC were estimated by their
H$\alpha$ luminosities without extinction correction.  If a part of
ionizing photons are absorbed by dust before hydrogen ionization,
their energy is converted into IR luminosity.  Therefore, the sum of
SFRs estimated by H$\alpha$ and IR luminosity gives a good upper bound
on the total SFR.  (It is an upper bound because non-ionizing photons
also contribute to IR luminosity when they absorbed by dust.)  The
SFRs estimated by IR luminosities of SMC and LMC are $0.0153 \pm
0.0046$ and $0.146 \pm 0.04 \ \rm M_\odot/yr$, respectively, by the
same method as for SFR estimates of M 82 and NGC 253 described in
\S\ref{sec:ModelingLg}.  These are 40--60\% of SFRs by H$\alpha$, 
indicating that these galaxies are not heavily obscured by dust.  
When we add the SFRs estimated by IR luminosity to
the SFRs of SMC and LMC, the EGRB flux from star-forming galaxies
changes slightly by a factor of 1.2.  On the other hand, starburst
galaxies (M82, NGC 253) have large extinctions and hence SFRs estimated
by IR luminosity are much bigger than those by H$\alpha$, as mentioned
in \S\ref{sec:ModelingLg}.

In the previous version of this paper put on the preprint server
(arXiv:1005.1390v1), we reported about 14\% contribution from
star-forming galaxies to the total EGRB flux, which is about two times
bigger than our new result.  We examined the reason for this, and
found that it is mainly due to the smaller SFR of LMC used in the
previous version, which was estimated only by IR luminosity.  As
mentioned above, SFR from H$\alpha$ luminosity is about two times
larger than that from IR for LMC ans SMC.  (Adding SMC to the nearby
sample is also a new point of the current version, but it does not
significantly affect the EGRB flux.) Therefore we believe that our
new result is more appropriate, but it would be better to consider
that the systematic uncertainty of the EGRB flux from star-forming
galaxies is not smaller than a factor of two, given the small number
of nearby galaxies with observed gamma-ray luminosities and
uncertainties in estimates of SFR and gas mass.

\begin{figure*}
  \begin{center}
    \begin{tabular}{cc}
      \resizebox{90mm}{!}{\includegraphics{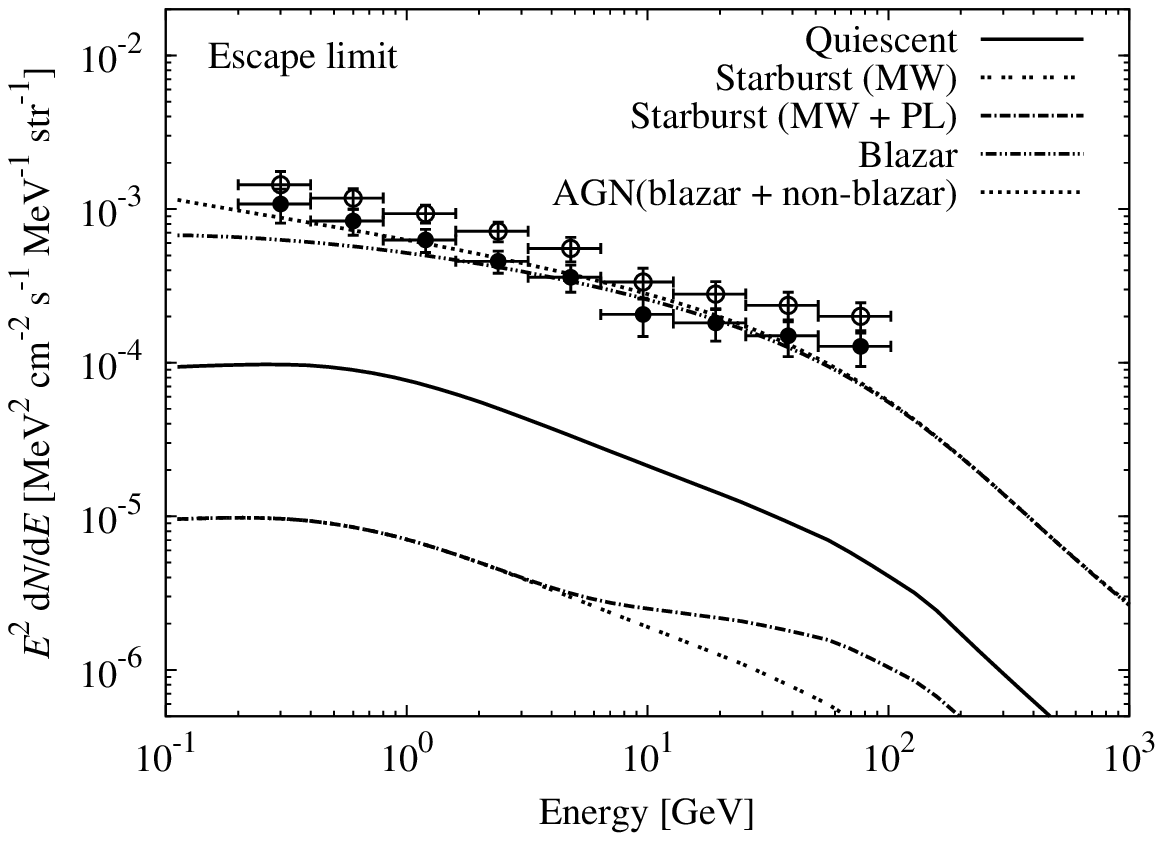}} &
      \resizebox{90mm}{!}{\includegraphics{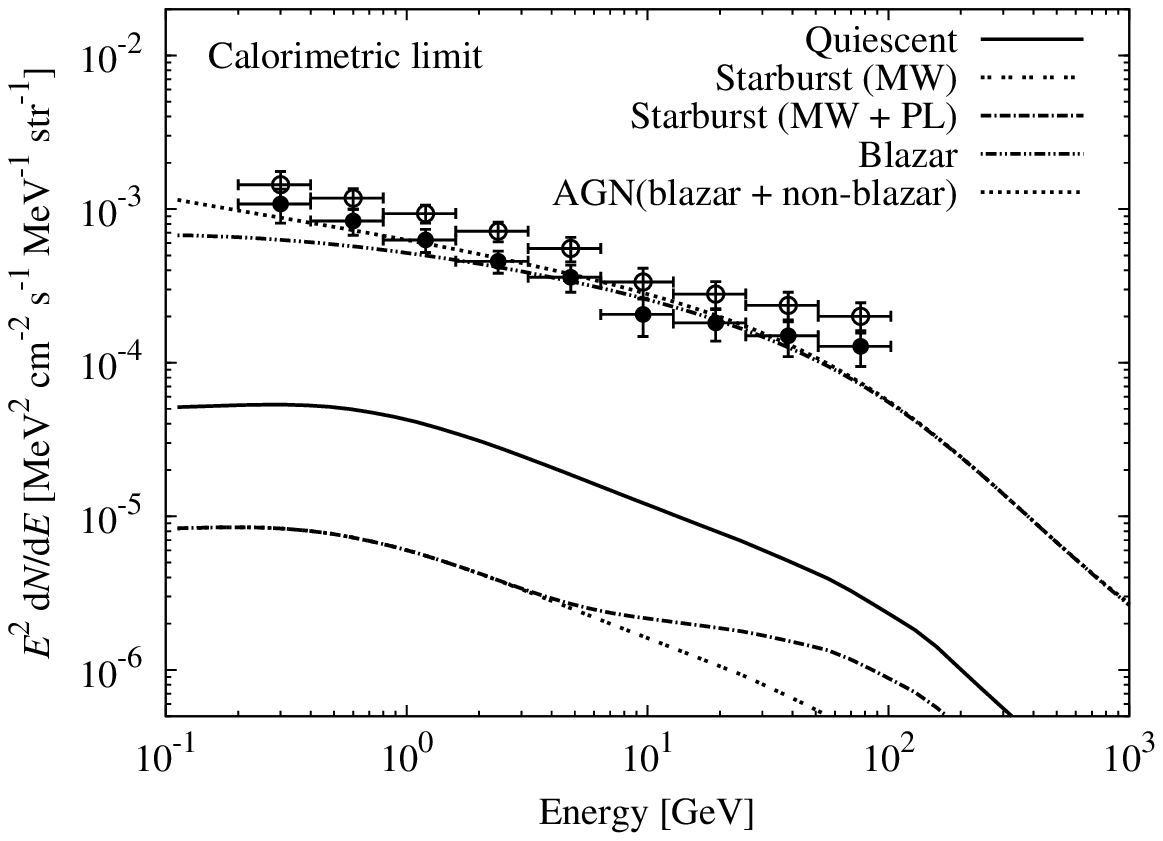}} \\\
    \end{tabular}
    \caption{ The EGRB spectrum. The theoretical predictions for the
      contributions from quiescent and starburst galaxies are
      separately shown, for the escape-limit (left panel) and the
      calorimetric limit (right).  For starburst galaxies, we plotted two cases
      corresponding to two different spectral shapes of MW and MW +
      power-law. For comparison, the EGRB models of blazars and all
      AGNs (blazar + non-blazar) by Inoue \& Totani (2009) are also
      shown.  Open circles show the total EGRB flux reported by {\it
        Fermi} (\citealt{PhysRevLett.104.101101}) including resolved
      sources, and the filled circles are the same but for the
      unresolved diffuse flux excluding resolved sources. The model
      curves should be compared with the open circle data.  }
\label{fig:egrb1}
\end{center}
\end{figure*}

\begin{figure*}
  \begin{center}
    \begin{tabular}{cc}\
	  \resizebox{90mm}{!}{\includegraphics{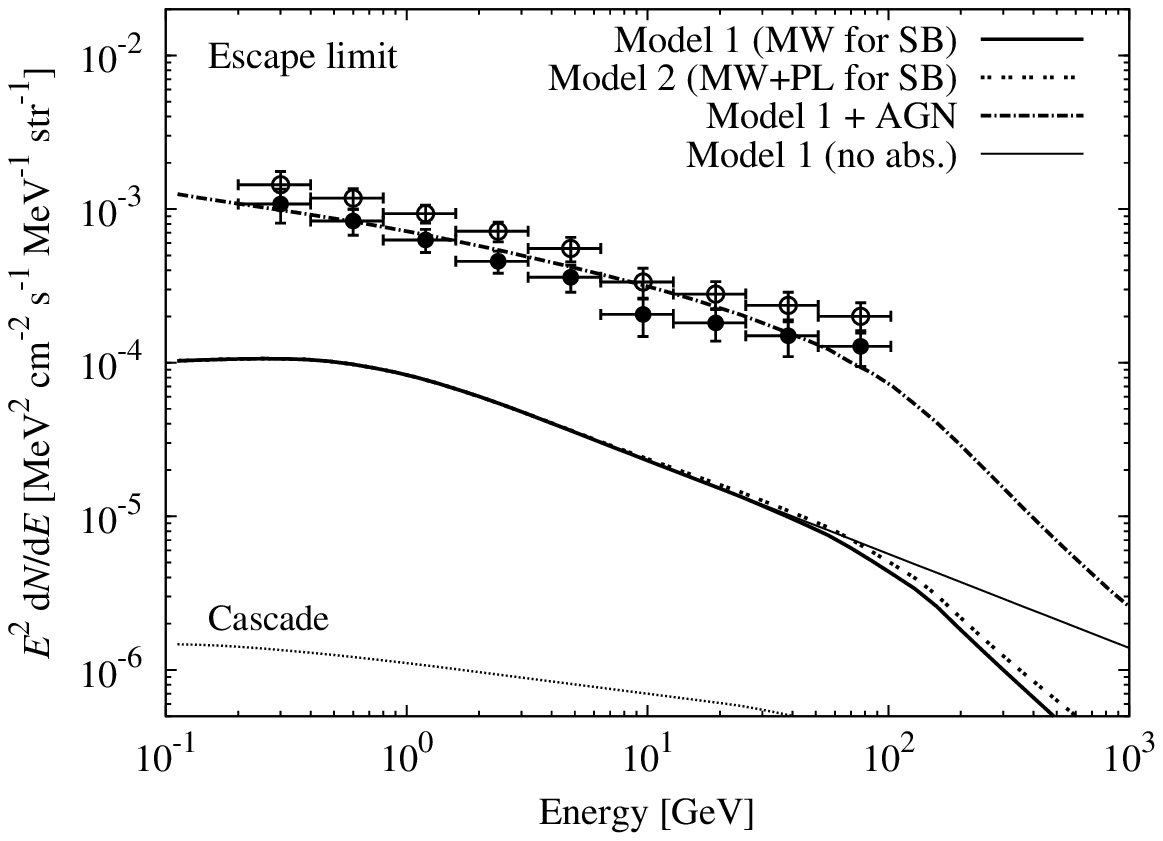}} &
      \resizebox{90mm}{!}{\includegraphics{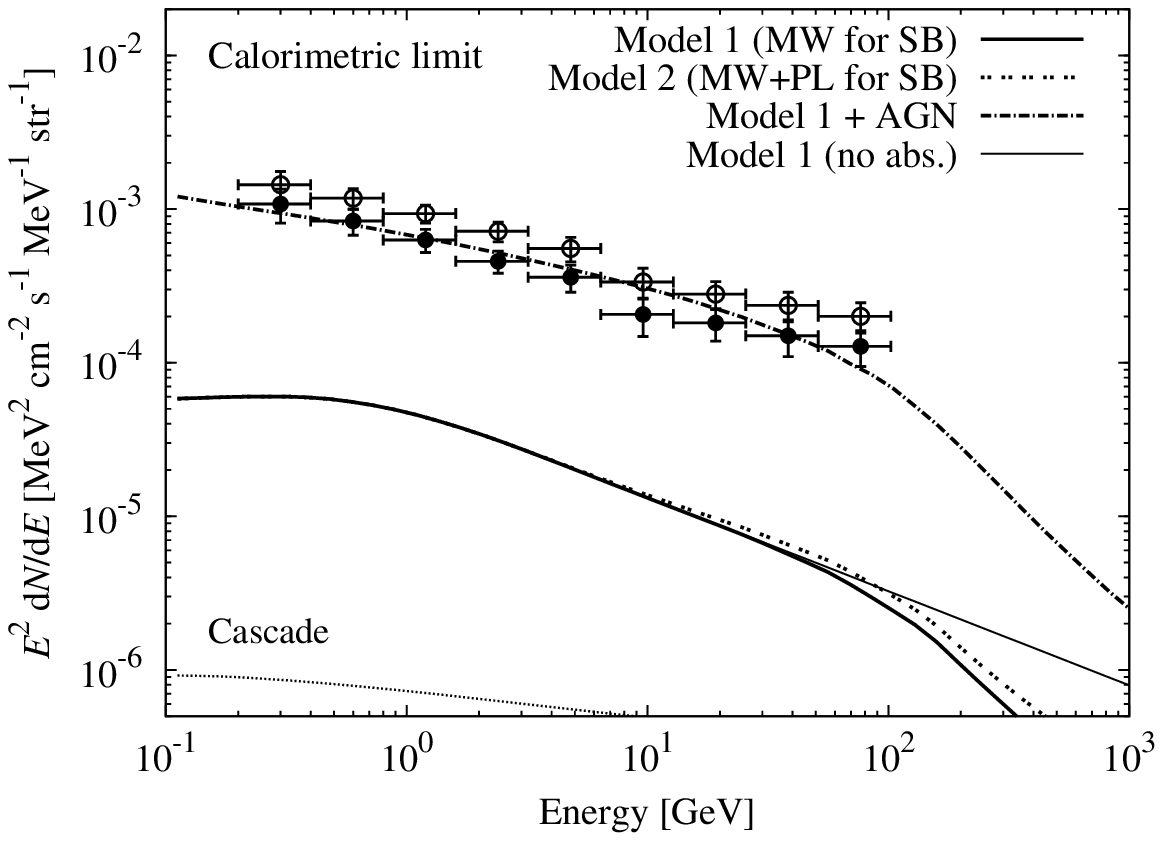}} \\
    \end{tabular}
    \caption{ The same as Fig. \ref{fig:egrb1}, but for the sums of
      each component are shown. The Models 1 and 2 are EGRB from all
      star-forming galaxies (quiescent plus starburst), with the two
      different spectra for starbursts.  To show the effect of
      intergalactic absorption, the Model 1 without absorption (thin
      solid line) and the cascade component of the Model 1 are also
      shown.  The total of star-forming galaxies plus AGNs is also
      shown using the Model 1 for star-forming galaxies.  }
\label{fig:egrb2}
\end{center}
\end{figure*}

\subsection{Implications for the Origin of EGRB}
\label{sec:EGRB-origin}

Our result indicates that about 10\% of EGRB can naturally be
explained by star-forming galaxies. Here we discuss the origin of EGRB
in general, also considering the larger contribution from AGNs. In
Figs. \ref{fig:egrb1} and \ref{fig:egrb2}, we plotted the EGRB
model of AGNs by Inoue \& Totani (2009, hereafter IT09).  The IT09
model includes two populations of AGNs: one is blazars, and the other
is non-blazars that are responsible for the EGRB at 1--10 MeV. Here
we plotted models of U03($q, \gamma_1$) and $\Gamma = 3.5$ (see
IT09 for details).

Many blazars have been detected by EGRET and Fermi, and it is obvious
that they are the major contributor to EGRB.  The blazar component of
IT09 can account for 43\% of the total EGRB flux, which is in good
agreement with the latest observational estimate of Abdo et
al. (2010d) based on the latest {\it Fermi} data\footnote{ Note that
  the number (16\%) quoted by Abdo et al. (2010d) is the fraction of
  unresolved blazars against the unresolved diffuse EGRB observed by
  {\it Fermi}. If this number is converted to that of all blazars
  against the total (unresolved + resolved components) EGRB flux, it
  increases to $\sim$40\% in good agreement with IT09.}.  Therefore,
adding the component from star-forming galaxies, more than 50\% of
EGRB can be explained by the sum of surely existing sources. The
non-blazar AGN component of IT09 has not yet been confirmed by
observations, but it is introduced to explain the background radiation
from hard X-ray to MeV region in a physically natural way (Inoue et
al. 2008). If this component extends to $\sim$ 100 MeV, the
contribution to EGRB flux at $>$100 MeV could be $\sim 20$\%, and
hence it is not difficult to explain $\gtrsim$ 70\% of EGRB by
theoretically reasonable sources.  It should also be noted that both
the AGN component and star-forming galaxy component have remarkably
similar spectra to the observed {\it Fermi} EGRB.

The $\lesssim$30\% residual of EGRB may not be explained simply by the
sum of AGNs and star-forming galaxies, and there may be a room for
different populations such as galaxy clusters or dark matter
annihilation. However, there are still uncertainties in the modeling
of AGN gamma-ray luminosity function.  The blazar luminosity function
of IT09 was based on EGRET data, including uncertainties about
detection and identification efficiencies depending on the location in
the sky. The recent observational estimate of the blazar contribution
by {\it Fermi} data includes a large correction about detection
efficiency near the sensitivity limit, which is generally dependent on
models. Blazars are highly variable sources, and it should add another
uncertainty in EGRB estimates. Therefore we conclude that a large
part of EGRB can be explained by known or physically reasonable
sources, and there is no strong evidence for exotic components like dark
matter annihilation.

\section{Conclusions}
\label{sec:Conclusions}

We have presented a new calculation of EGRB from cosmic-ray
interactions in star-forming galaxies, based on a state-of-the-art
galaxy formation model in the framework of hierarchical structure
formation. The galaxy formation model is quantitatively consistent
with various observations at high redshifts as well as
the local universe. Gamma-ray luminosities of galaxies are calculated by
star formation rate and gas mass in model galaxies, with the relation
$L_\gamma \propto (\psi M_{\rm gas})^{0.86}$ (the escape limit)
or  $L_\gamma \propto (\psi)^{1.2}$
(the calorimetric limit) which are calibrated by the recent
observational data of nearby galaxies.  The predicted number of nearby
galaxies that are detectable by {\it Fermi} is consistent with the
actual number observed so far.

We found that star-forming galaxies make 4--7\%
contribution to the total EGRB flux reported by {\it Fermi} in our
standard model. Combined with the contribution from blazars as
estimated by the {\it Fermi} data, more than $\sim$50\% of EGRB can be
accounted for. If the soft power-law tail of CXB is extending from MeV
to GeV region as expected from the MeV background data and
theoretical considerations of AGN accretion disks
(Inoue et al. 2008), additional $\sim$20
\% can be explained.  The combined spectrum by AGNs and star-forming
galaxies is remarkably similar to the observed EGRB spectrum.  It
should be noted that there is no free parameters that can be tuned to
fit the obeserved EGRB spectrum in the present model; the blazar
EGRB spectrum of IT09 is determined by the spectral templates of the
blazar SED (spectral energy distribution) sequence, and that of the
star-forming galaxy component in the present work by templates
constructed by observed nearby galaxies. Therefore we conclude that a
large part ($\gtrsim$70\%) of EGRB can be explained by reasonable
sources of AGNs and star-forming galaxies.  Further examination is
required to see whether the apparent $\lesssim$30\% residual of EGRB is
mainly a result of modeling uncertainties, experimental/observational
uncertainties in deriving the EGRB data, or significant contributions
from completely different source populations.  Given the good spectral
agreement of the {\it Fermi} data and AGNs/star-forming galaxies, the rest
of EGRB is also likely dominated by astrophysical sources accelerating
particles, even if a completely diffrent population is responsible for
it.

It is in contrast that a large part of the blazar component of EGRB
will be resolved into discrete sources by the ultimate {\it Fermi}
sensitivity in the near future (IT09), while almost all of the
star-forming galaxy component will remain unresolved because of the
faintness of individual sources. Therefore, any exotic contribution
like dark matter annihilation cannot be probed directly under the
level of the star-forming galaxies, i.e., $\sim$ 10\% of the total
EGRB flux.  Another approach such as utilizing anisotropy would be
required to search for the signal under that level (see, e.g. Ando et
al. 2007; \citealt{2009PhRvL.102x1301S},
\citealt{2009arXiv0912.1854H}, \citealt{2010arXiv1005.0843C})

We assumed a simple empirical relation between gamma-ray luminosity,
SFR, and gas mass of galaxies.  Only two spectral templates were used
in the calculation. A next step would be to construct a more physical
model of gamma-ray luminosity and spectrum based on a larger number of
physical quantities (e.g., size and stellar radiation field in
addition to SFR and gas mass), taking into account propagation of
cosmic-rays and production processes of gamma-rays.

\acknowledgments 
We would like to thank the anonymous referee for many
useful comments. 
This work was supported by the Grant-Aid for the
Global COE Program "The Next Generation of Physics, Spun from
University and Emergence " from the Ministry of Education, Culture,
Sports, Science and Technology (MEXT) of Japan.  The numerical
calculations were in part carried out on SGI Altix3700 BX2 at Yukawa
Institute for Theoretical Physics of Kyoto University.  
TT was supported by the Grant-in-Aid for Scientific Research
(19047003, 19740099, 2004005) from MEXT.  MARK was supported by the Research
Fellowship for Young Scientists from the Japan Society for the
Promotion of Science (JSPS).

\end{document}